\documentclass[conference]{IEEEtran}
 \IEEEoverridecommandlockouts
\usepackage{cite}
\usepackage{amsmath,amssymb,amsfonts}
\usepackage{graphicx}
\usepackage{textcomp}
\usepackage{xcolor}
\usepackage{comment}
\usepackage{optidef}
\usepackage[T1]{fontenc}
\usepackage{mathtools} 
\usepackage{cuted}
\usepackage{flushend}
\usepackage{bbm}
\usepackage{dsfont}
\usepackage{mathtools}
\usepackage{multirow}
\usepackage{float}
\usepackage{dirtytalk}
\usepackage{array}
\usepackage{hyperref}

\begin{document}
\title{Decomposition Based Interference Management Framework for Local 6G Networks
\thanks{The research leading to this paper was supported by the Research Council of Finland (former Academy of Finland) \href{https://www.6gflagship.com/}{6G Flagship program} (Grant Number: 346208), and Business Finland's 6GBridge program through the projects Local 6G (Grant Number 8002/31/2022) and 6CORE (Grant Number 8410/31/2022).}}
 \author{\IEEEauthorblockN{Samitha~Gunarathne, Thushan~Sivalingam, Nurul~Huda~Mahmood, Nandana~Rajatheva, and Matti~Latva-Aho}
 \IEEEauthorblockA{6G Flagship, Centre for Wireless Communications, University of Oulu, Oulu, Finland}
\{samitha.gunarathne, thushan.sivalingam, nurulhuda.mahmood, nandana.rajatheva, matti.latva-aho\}@oulu.fi
 }
 \maketitle
\begin{abstract}
Managing inter-cell interference is among the major challenges in a wireless network, more so when strict quality of service needs to be guaranteed such as in ultra-reliable low latency communications (URLLC) applications. This study introduces a novel intelligent interference management framework for a local 6G network that allocates resources based on interference prediction. The proposed algorithm involves an advanced signal pre-processing technique known as empirical mode decomposition followed by prediction of each decomposed component using the sequence-to-one transformer algorithm. The predicted interference power is then used to estimate future signal-to-interference plus noise ratio, and subsequently allocate resources to guarantee the high reliability required by URLLC applications. Finally, an interference cancellation scheme is explored based on the predicted interference signal with the transformer model. The proposed sequence-to-one transformer model exhibits its robustness for interference prediction. The proposed scheme is numerically evaluated against two baseline algorithms, and is found that the root mean squared error is reduced by up to $55\%$ over a baseline scheme. 
\end{abstract}
\vspace{3mm}
\begin{IEEEkeywords}
Deep neural networks, empirical mode decomposition, transformers, URLLC.
\end{IEEEkeywords}

\section{Introduction}
% 1st paragraph 
%(Introduction about 5G and 6G service categories) 
With the advancement of technology, the demand for wireless communication has been growing exponentially over the past few years. 
According to the statistics, the calculated global mobile traffic volume was around $7.5$ EB/month in 2010 \cite{chowdhury20206g}. It is forecasted to reach 5016 EB/month by 2030, along with other applications requiring network capabilities beyond that which can be supported by 5G \cite{chowdhury20206g, 7982949}. Research is being carried out to introduce the sixth generation (6G) wireless network to fulfill this capacity.

% 2nd Paragraph 
%(requirements of 6G, emphasis about URLLC) 
The current fifth generation (5G) New Radio (NR) introduced three distinct service classes to efficiently support the huge and diverse traffic demand. One of these service classes is ultra-reliable low latency communications (URLLC) targeting vertical areas such as industrial automation, intelligent transportation systems and smart healthcare. URLLC demands low error rates in the order of $10^{-5}$ along with sub millisecond latency. %6G will be developed for ubiquitous mobile ultra-broadband (uMUB), ultra-high speed with low latency communications (uHSLLC), massive machine-type communication (mMTC), and ultra-high data density (uHDD) \cite{chowdhury20206g}. 
In order to guarantee ultra-low end-to-end transmission latency, URLLC traffic is generally assumed to have short packet sizes. %For 6G, this E2E delay is required as at most 0.1 ms while the peak data rate for 6G uMUB is expected to provide up to 1 Tbps \cite{8636206}. 

Efficient management of inter-cell interference is among the major challenges in guaranteeing the stringent reliability and latency requirement of URLLC. Conventional approaches such as hybrid automatic repeat request (HARQ) that are rather efficient for enhanced mobile broadband (eMBB) applications are not well suited for low latency applications. Instead solutions that can predict the interference conditions and allocate resources proactively are found to be more effective~\cite{12}. 

% 5th Paragraph
%(mMTC Intro)
%In 6G, mMTC is one of the leading service requirements; since the exponential growth of intelligent devices and applications, several challenging tasks exist to achieve its requirements. mMTC has a wide range of applications, such as factory automation, autonomous driving, smart healthcare, sensing and smart metering, robotics, logistic applications, and massive IoT (mIoT) \cite{dogra2020survey}. However, according to each application, the quality of service (QoS) requirement differs. Even though the QoS is different, after establishing the connection between two or more devices, reliability, low latency, and security are instrumental to communication \cite{dogra2020survey}.
% \subsection{Existing Work}

% 7th Paragraph 
%(Related work)
The work in \cite{11} provides information about interference prediction in wireless networks using a non-linear auto-regressive neural network (NARNN) for URLLC purposes. The proposed method of \cite{11} achieved a reasonable interference prediction accuracy. Reference \cite{12} shows a novel interference forecasting method by modeling the variation of interference as a discrete state space discrete-time Markov chain. The specialty of this algorithm is it considers the entire interference distribution. The authors of \cite{22} proposed an interference prediction method for beyond 5G (B5G) networks to increase the URLLC and link adaptation (LA). A kernel-based probability density estimation algorithm is used for this prediction. The work in \cite{21} presents a deep learning-based interference prediction graph for moving mIoT small cells. The deep learning algorithm used for this study is long short-term memory (LSTM) network, and the interference prediction output is used for resource allocation purposes. The work in \cite{jayawardhana2023predictive} is closely related to the objective of our research. In~\cite{jayawardhana2023predictive}, wireless interference has been predicted using LSTM and auto-regressive integrated moving average (ARIMA) algorithms. As a pre-processing method, empirical mode decomposition (EMD) is used before the prediction models. Based on the literature, there is room to further investigate intelligent interference prediction solutions that lead to efficient interference management. 

The main contribution of this study is to propose and implement a novel, intelligent interference management scheme specifically targeting 6G local applications. A local 6G network is referred to as a network covering a limited area with connected multiple devices, which improves the operating efficiency for specific vertical sectors such as industries serving URLLC applications. %For example, hospitals, factories, shopping malls, universities, etc. 
Our proposed scheme is based on a deep neural network (DNN) algorithm, namely transformers \cite{32}, and works as follows. First, the aggregate interference signal is decomposed into multiple signals called intrinsic mode functions (IMF) via EMD. Different IMFs have different principle frequencies, making them easier to predict using DNNs compared to predicting the composite signal. A transformer model is then trained to predict each IMFs individually leading to the predicted interference signal composed of the sum of the predicted IMFs. Usually, transformers are used for natural language processing (NLP) tasks such as the ability to understand text and spoken words by computers. But in this study, we employed transformers as a univariate time series forecasting method. Finally, the predicted interference signal is used as an input to different resource allocation algorithms.

% Since wireless interference contains signals with numerous frequencies, the EMD algorithm is used as a signal pre-processing technique for decomposing. Usually, transformers are used for natural language processing (NLP) tasks such as the ability to understand text and spoken words by computers. But in this study, we employed transformers as a univariate time series forecasting method. 

\begin{comment}
Add here one paragraph with a summary of a few current literature and how they approached this problem of interference prediction. (some examples: https://ieeexplore.ieee.org/document/9354031 
https://ieeexplore.ieee.org/document/9759361
https://ieeexplore.ieee.org/document/9428512)
Then say that: 
Since interference is a highly random signal with contributions from multiple interferers, predicting it accurately is a complex procedure. Hence, more accurate prediction mechanisms need to be explored. Towards this end, we propose a novel empirical mode decomposition (EMD) based approach.
\end{comment}

\section{System Model}
\label{sec:systemModel}
\begin{figure}[tb]
    \center
    \includegraphics[width=0.9\linewidth]{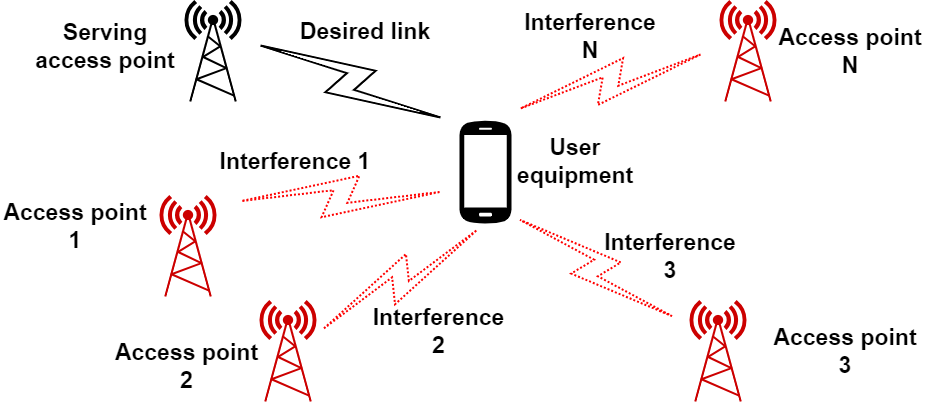}
    \caption{System model.}
    \label{fig:system_model}
\end{figure}

The considered local 6G network is shown in Fig. \ref{fig:system_model}. The primary focus of this work is on the downlink channel of a local 6G wireless network with $N$ number of interfering links. In order to establish a clear and strong research objective, we will begin by outlining the essential assumptions as follows. 

We target a local 6G wireless communication network, where a desired user is served by a serving access point in the presence of interference from other access points operating in the same frequency. In wireless communication, mobility and dynamic behavior are known facts. But since we are mainly considering indoor scenarios, the mobility is negligible. Therefore, we assume the environment is static. We assume the communication channels have multipath propagation and reflections to model the interference. Therefore, this scenario is statistically modeled by Rayleigh fading. Assume the desired link has an average signal-to-noise ratio (SNR) of $\bar{\gamma}$. Also, the maximum and minimum mean interference-to-noise ratio (INR) values of interference are denoted by $\gamma_{max}$ and $\gamma_{min}$, respectively. Similar to the SNR, the mean INR represents the mean power of each interferer signal. We assume that the desired receiver is served by the strongest access point, which implies that $\bar{\gamma}>\gamma_{max}$. Finally, we assume no cooperation between transmitters to manage the interference collaboratively.  
\begin{comment}
In this study, the downlink of a wireless network is examined. The objective is to determine the optimal resource allocation for a URLLC service operating in the presence of $N$ interferers, as shown in Fig. \ref{fig:system_model}. It is assumed that the desired channel has a signal-to-noise ratio (SNR) of $\bar{\gamma}$ and the interference-to-noise ratios (INR) for each interfering link are uniformly distributed within a given range $[{\gamma}_{min}, {\gamma}_{max}]$. The above mentioned INR can also be referred to as the SNR of an interference signal.
\end{comment}

\section{Methodology}
\label{sec:methodology}
\begin{comment} A schematic of the proposed prediction method is shown in Fig. \ref{fig:predict_mode}. 
\begin{figure}[t]
    \center
    \includegraphics[width=\linewidth]{each_block.png}
    \caption{Proposed method for interference prediction.}
    \label{fig:predict_mode}
\end{figure}
EMD based hybrid prediction method is used with two prediction methods LSTM and ARIMA. Resources are efficiently allocated using finite block length theory relying on the predicted interference values. The total interference signal power that the mobile station experiences is derived by adding all the individual interference signal powers as
\begin{equation}
\label{eqn:total_interference}
{f}(t) =\sum_{i=1}^{N} {Interference}{(i)},
\end{equation}
where ${Interference}{(i)}$ is the interference signal power from the $i$\textsuperscript{th} interferer and ${f}(t)$ is the total interference power observed by the UE.
\end{comment}

According to the system model depicted in Fig.\ref{fig:system_model}, and considering the problem statement, we need to forecast the wireless interference affecting the user equipment (UE). The UE experiences the total interference signal as a mixture or combination of interference from each interferer. Therefore, the received signal at the UE can be expressed as
\begin{equation}
    \label{eqn:received_signal}
    R_s(t) = \sqrt{E_s}h_s(t)s_0(t) + \sum_{i = 1}^{N} \sqrt{E_i}h_i(t) s_i(t) + w(t),
\end{equation} 
where $R_s(t)$ is the received signal, $s_0(t)$ is desired signal, $s_i(t)$ is $i$-th interference signal,  $E_s$ is desired signal power, $E_i$ is interfered signal power, $h_s(t)$ is desired transmitted channel response, $h_i(t)$ is $i$-th interfered channel response, and $w(t)$ is additive white Gaussian noise (AWGN).
The objective is to predict future interference using past interference data. 
To forecast interference in this study, we employ two different approaches as follows:

\begin{enumerate}
    \item{Time domain conventional approach}
    \item{Time domain proposed approach}
\end{enumerate}
Under each approach, we utilized transformers as our main prediction algorithm. For comparison purposes, we applied one deep learning algorithm, specifically LSTM, and one statistical signal processing algorithm, ARIMA, by referring to research work \cite{jayawardhana2023predictive}. Moreover, we incorporated EMD as a pre-processing technique. Finally, the results are compared to choose the best algorithm.

\subsection{Empirical Mode Decomposition} \label{emd}
EMD \cite{34} is a powerful signal processing technique for analyzing time-frequency signals. EMD is capable of decomposing a signal into several sub-signals called intrinsic mode functions (IMFs) by referring to its local oscillatory characteristics. These intrinsic modes can be considered as bases of the original signal, and this decomposition is taken place by the EMD algorithm itself without defining any pre-deterministic modes \cite{34}. In this study, we utilized EMD to decompose the interference signal into multiple IMFs with the aim of improving the prediction accuracy.

Usually, after decomposing a signal using EMD, the decomposed IMFs are frequency-ordered. The leftover component is called the residual. %In this research, the power of the interference signal is decomposed into IMFs and a residual as a signal pre-processing technique. 
Therefore, the $i^{th}$ interference signal power denoted as $I_i(t) = \sqrt{E_i}h_i(t) s_i(t)$ can be represented as a summation of a set of IMFs with different frequencies and the residual signal as
\begin{equation}
    \label{eqn:decomposed_signal}
    I_i(t) =  \sum_{j = 1}^{J_i}IMF_j(t) + res(t),
\end{equation} 
where $IMF_j(t)$ is $i$-th IMF and $res(t)$ is the residual signal. The number of IMFs, $J_i$, depends on the frequency components of $I_i(t)$. Most mathematical software packages like Matlab and Mathematica have built in functions for EMD.

\subsection{Transformers}

Transformers is a deep learning architecture that can transform a sequence into another sequence with the help of its state-of-the-art of attention mechanism. Transformer architecture was initially introduced in 2017 by a research group in Google in \cite{32}. After introducing transformers, it has been applied to many real-world applications such as NLP, computer vision, conversational chatbots, etc \cite{39, 40}. Due to its superior characteristics, it has significantly outperformed most of the previous deep learning approaches, namely NLP and recurrent neural networks (RNNs) in computer vision.

\begin{figure}[tb]
\center
\includegraphics*[width= 0.25\textwidth]{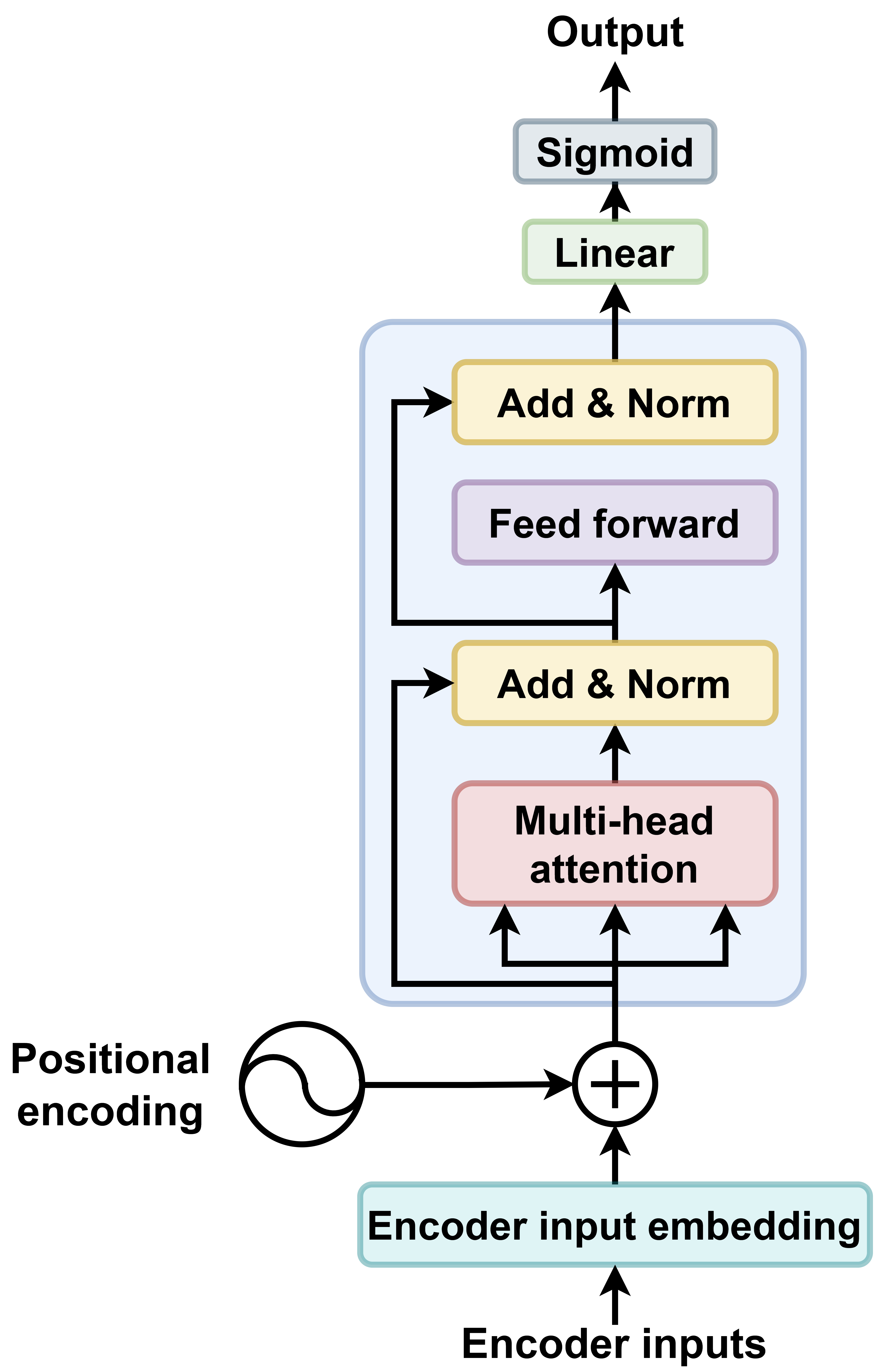}
\caption{Sequence-to-one transformer model architecture.}
\label{fig:TF_architecture_s21}
\end{figure}

In general, RNNs have the issue of vanishing gradients. LSTM was introduced to resolve this issue; however, the LSTM model's memory can only deal with a certain length of long memory while LSTM is slow in training because its input is fed sequentially. Thanks to the attention mechanism, transformers gain the ability to have extremely long data sequence memory. Sequence-to-sequence transformer model architecture mainly consists of two parts such as encoder and decoder. However, in this research, the transformer model was adapted into time series interference forecasting. Therefore, we modified the initial sequence-to-sequence transformer model into the sequence-to-one model, which only contains an encoder layer. Fig. \ref{fig:TF_architecture_s21} illustrates the modified transformer architecture for sequence-to-one prediction.

\subsection{Prediction Approach and Training}
In the context of this study, which is mainly associated with time series data, essentially needs to be determined the appropriate length of data for training and validation. According to the signal behaviour, a reasonable $80\%$ of input interference data samples are allocated for training while the remaining $20\%$ of samples are allocated for validating. Since this research is dealt with wireless communication interference, it is not good to take excessively long past data sequence and overly too short data sequences. In this study, we primarily employed two distinct approaches, which are outlined below.

% In the context of this study, which is mainly associated with time series data, essentially needs to be determined the appropriate length of data for training and validation. According to the signal behaviour, a reasonable $80\%$ of input interference data samples are allocated for training while the remaining $20\%$ of samples are allocated for validating. Since this research is dealt with wireless communication interference,  the length of interference data fed to the algorithms for training is essentially important. The main reason for this is, wireless interference exhibits random characteristics in nature and therefore, it is not good to take excessively long past data sequence and overly too short data sequences. In this study, we primarily employed two distinct approaches, which are outlined below.

 \subsubsection{Time domain proposed approach}
\begin{figure*}[tb]
    \center
    \includegraphics[width=0.75\linewidth]{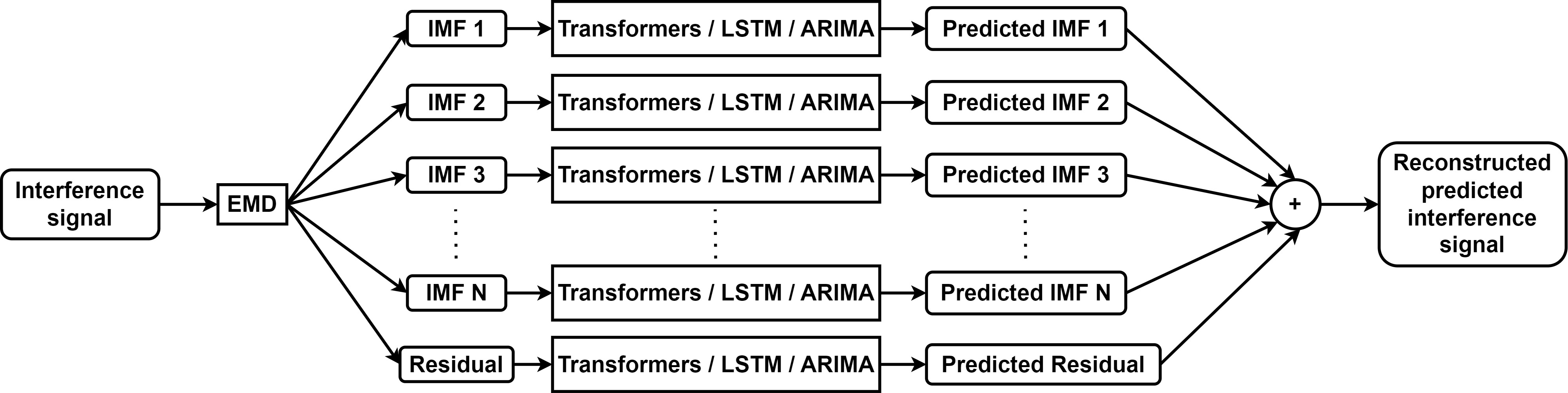}
    \caption{Time domain proposed interference forecasting scheme.}
    \label{fig:proposed_method}
\end{figure*}

One drawback of the conventional approach is the degree of randomness of the interference signal is high. This randomness is a challenge for the above prediction algorithms as they are unable to capture the temporal or statistical features of the past total interference signal due to its randomness. As explained in Section \ref{emd}, EMD is used as an advanced signal pre-processing technique to decompose interference signals. After decomposing, EMD outputs $K$ number of IMFs and one residual. Then apply, transformers, LSTM, and ARIMA algorithms into the set of IMFs and residuals separately to train the algorithms. These algorithms predict each IMF and residual, and finally, the predicted interference signal using transformers is reconstructed as
\begin{equation}
    \label{eqn:reconstruction_TF}
    I_i(t)_{TF} = \sum_{i = 1}^{K} IMF_{TF}^{(i)}(t) + R_{TF}(t),
\end{equation} 
where $I(t)_{TF}$, is the reconstructed interference signal obtained after the prediction using the transformer algorithm. $IMF_{TF}$ is the predicted IMFs of the transformer algorithm. Also, $R_{TF}(t)$is the predicted residual component of the transformer model. The same mathematical representation can also be used for LSTM and ARIMA models. Fig. \ref{fig:proposed_method} illustrates the complete Time domain proposed interference forecasting scheme.

 \subsubsection{Time domain conventional approach}
Under the time domain conventional approach, the process involves applying transformers, LSTM, and ARIMA directly to the original interference signal without decompose by EMD and observing the forecasted interference signal. Here, the original interference signal is fed to each algorithm separately. %Figure \ref{fig:conventional_method} illustrates the conventional approach using a flow diagram.

\subsection{Resource Allocation}
\label{Resource_allocation}
In order to formulate the resource allocation mathematically, let's assume $P_s$ is the desired signal which is known from channel state information (CSI) \cite{jayawardhana2023predictive}. Therefore, the predicted SINR $\hat{\delta}$ can be written as \cite{38}
\begin{equation}
    \label{eqn:pred_SINR}
    \hat{\delta} = \frac{P_s}{\hat{I_p} + N_0},
\end{equation} 
where the normalized noise power is denoted by $N_0$ and the interference power that is predicted is denoted by $\hat{I_p}$. 
The block error rate \cite{jayawardhana2023predictive} can be approximated
\begin{equation}
    \label{eqn:decoding_error}
    \varepsilon \approx{Q\left(\frac{C(\hat{\delta} )R-D}{\sqrt{V(\hat{\delta} )R}}\right)}.
\end{equation} 
where $D$ is the number of maximum achievable information bits, $R$ is the number of channels in an AWGN channel, and $\varepsilon$ is the given decoding error probability. $C(\hat{\delta})$ is the Shannon capacity of AWGN channel and $V(\hat{\delta})$ is the dispersion of the channel.

Note that the resource allocation occurs a time instant prior to the actual transmission.  After transmission of $D$ bits, compute the actual SINR denoted by $\delta$ using the equation
    
\begin{equation}
    \label{eqn:act_SINR}
    \delta = \frac{P}{I + N_0},
\end{equation} 
where $I$ and $P$ are actual total interference and actual signal power respectively.
Therefore, the achieved error probability $\bar{\varepsilon}$ with the actual SINR $\delta$ using equation
\begin{equation}
    \label{eqn:achieved_error}
    \bar{\varepsilon} \approx {Q\left(\frac{C(\delta)R-D}{\sqrt{V(\delta)R}}\right)}.
\end{equation} 
Finally, the target and achieved error probabilities are plotted in the same graph.
In order to validate the performance of resource allocation, two baseline schemes were used for comparison.

\subsubsection{Genie-aided estimation: } This estimation scheme is considered a perfect estimator because it has complete knowledge of inference conditions before the transmitter. Therefore, this estimation scheme is considered an optimal baseline estimator, even though this behavior is not practical in the real-world environment \cite{12}.
\subsubsection{Moving average based estimation: } This estimator is originally adopted as LA for eMBB services. However, MA based estimator is a weighted average estimator, which can be applied to the interference estimation as \cite{12}

\begin{equation}
    \label{eqn:MA_estimator}
    \hat{J}_{t+1} = \alpha J_{t-1} + (1 - \alpha)\hat{J_t}, \quad\quad 0 < \alpha < 1
\end{equation} 
where $\alpha$ is the forgetting factor and $\hat{J_t}$ is the infinite impulse response (IIR) of the measured interference at time $t$ \cite{12}.

\subsection{Interference Cancellation}
\label{Interference_cancellation}
For the interference cancellation task, we only considered the proposed approach with transformers. Let $I(t) = \sum_{i = 1}^{N} \sqrt{E_i}h_i(t) s_i(t)$ denote the interference signal. Herein we assume that the interferer message $s_i(t)$ is correlated across time making $I(t)$ amenable to prediction. The interference signal consists of two components, namely the real and the quadrature part. First, both real and quadrature parts are forecasted separately using the proposed transformer model. %Therefore, $\Re(I(t))$ and $\Im(I(t))$ are considered real and quadrature components, respectively. 
Let us denote the predicted real and quadrature components are denoted as $\hat{\Re(I(t))}$ and $\hat{\Im(I(t))}$. Therefore, interference cancellation can be mathematically defined as
\begin{equation}
    \label{eqn:real_cancellation}
    \bar{\Re(I(t))} = \Re(I(t)) - \hat{\Re(I(t))},
\end{equation} 
\begin{equation}
    \label{eqn:im_cancellation}
    \bar{\Im(I(t))} = \Im(I(t)) - \hat{\Im(I(t))}.
\end{equation}

\section{Simulations Results}
\label{sec:sim_results}
Prior to proceeding with the prediction part, initially, both the desired and the interference signals were generated according to Table \ref{tab1}. Also, all the parameters related to resource allocation are summarized in the same table.

\begin{table}[htbp]
\caption{All the simulation parameters.}
\begin{center}
\begin{tabular}{l l}
\hline
\textbf{Parameter}&\textbf{Value} \\
\hline
SINR value of desired signal ($\bar{\gamma}$) & 20 dB \\
Number of interfered signals & 6 \\
INRs of interferes (dB)& 5, 2, 0, -3, -10, 1 \\
Number of samples considered & 200 for transformers \\
 & 100 for LSTM and ARIMA \\
Channel model & Rayleigh block fading \\
Target error rates ($\varepsilon$) & $10^{-5},10^{-4},10^{-3},10^{-2},10^{-1}$ \\
Forgetting factor ($\alpha$) & 0.01\\
Number of bits for resource & \multirow{2}{*}{50}\\
allocation ($D$) & \\
\hline
\end{tabular}
\label{tab1}
\end{center}
\end{table}

\subsection{Interference Power Prediction Results and Analysis}
We mainly focused on transformers in this work, while LSTM and ARIMA were used for comparison purposes. All the hyper-parameters used for the training phase related to the transformer model are tabulated in Table \ref{tab2}.

\begin{table}[htbp]
\caption{Hyper-parameters used in transformer model.}
\begin{center}
\begin{tabular}{l l l}
\hline
\textbf{Algorithm}&\textbf{Hyper-parameter}&\textbf{Value} \\
\hline
\multirow{10}{*}{Transformers} & Training window & 10 \\
& Number of Transformer blocks & 16 \\
& Attention head size & 32 \\
& Number of attention heads & 16 \\
& Feed forward dimension & 4 \\
& Dropout & 0.2 \\
& Loss function & MSE \\
& Activation function & ReLU \\
& Optimizer & Adam\\
& Number of epochs & 100 \\
\hline
\end{tabular}
\label{tab2}
\end{center}
\end{table}

For both LSTM and ARIMA models, the training window and number of epochs were taken as 30 and 100, respectively. LSTM architecture contains three LSTM layers with 16 neurons each, followed by two dense layers with sizes of 8 and 1. The loss function for the LSTM model is MSE, the activation function is ReLU, and the optimizer is Adam. In the ARIMA model, the order of the $AR(p)$, $MA(q)$ and integrate are $p = 30$, $q = 0$, and $d = 1$ respectively.

Prediction performance evaluation is a crucial measure to assess the effectiveness of the algorithms. There are numerous performance evaluation criteria for traditional statistical prediction methods and deep neural network algorithms according to their application. Since the task of this study is time series forecasting, root mean square error (RMSE) has been taken as the prediction evaluation criterion. It measures the RMSE between the predicted interference signal and the actual values in the validation dataset.

\begin{figure}[htbp]
    \center
    \includegraphics*[width=0.85\linewidth]{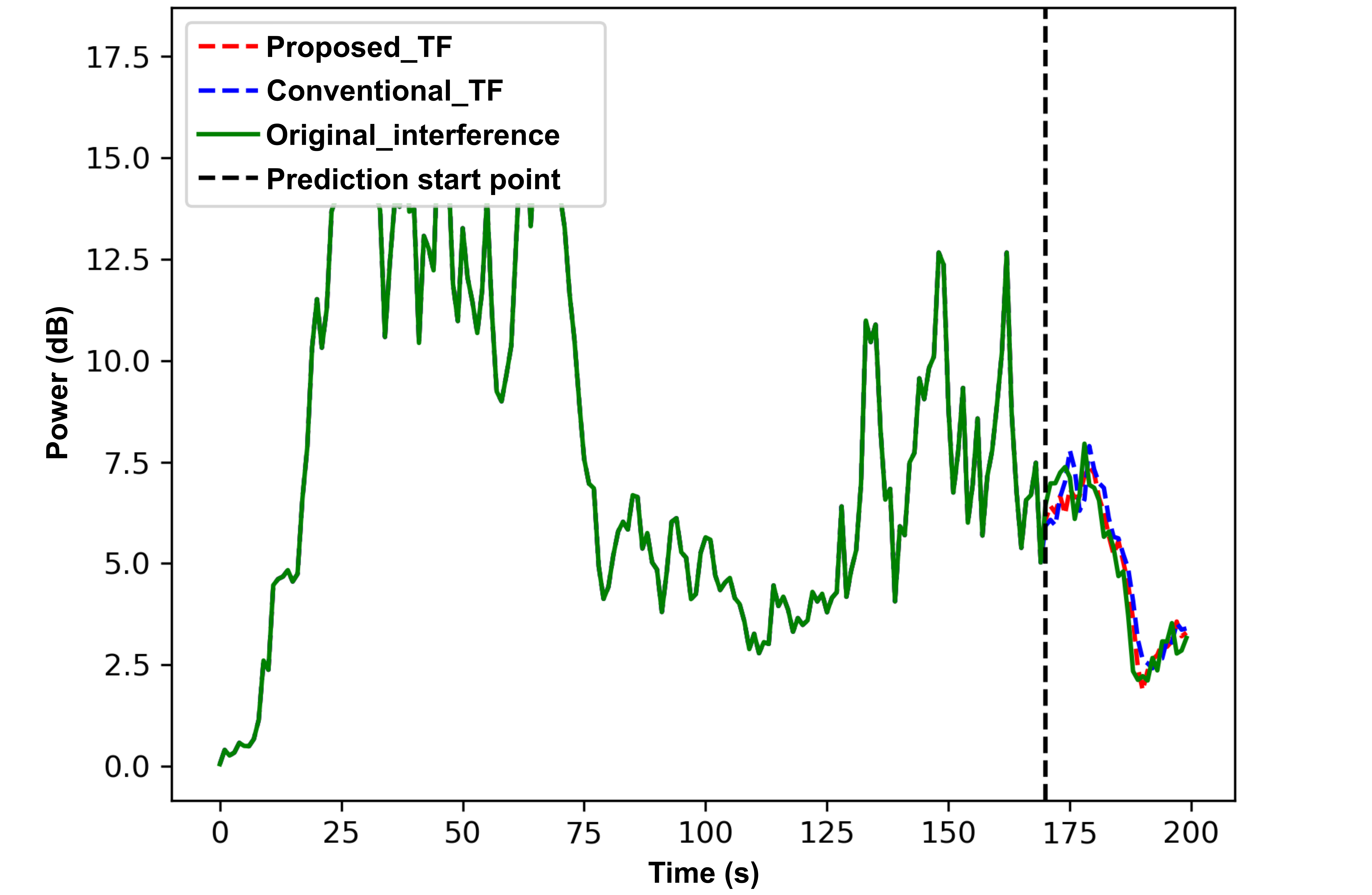}
    \caption{Comparison between proposed and conventional methods using transformer model.}
    \label{fig:TF_comparison}
\end{figure}

\begin{figure}[ht]
    \center
    \includegraphics*[width=0.85\linewidth]{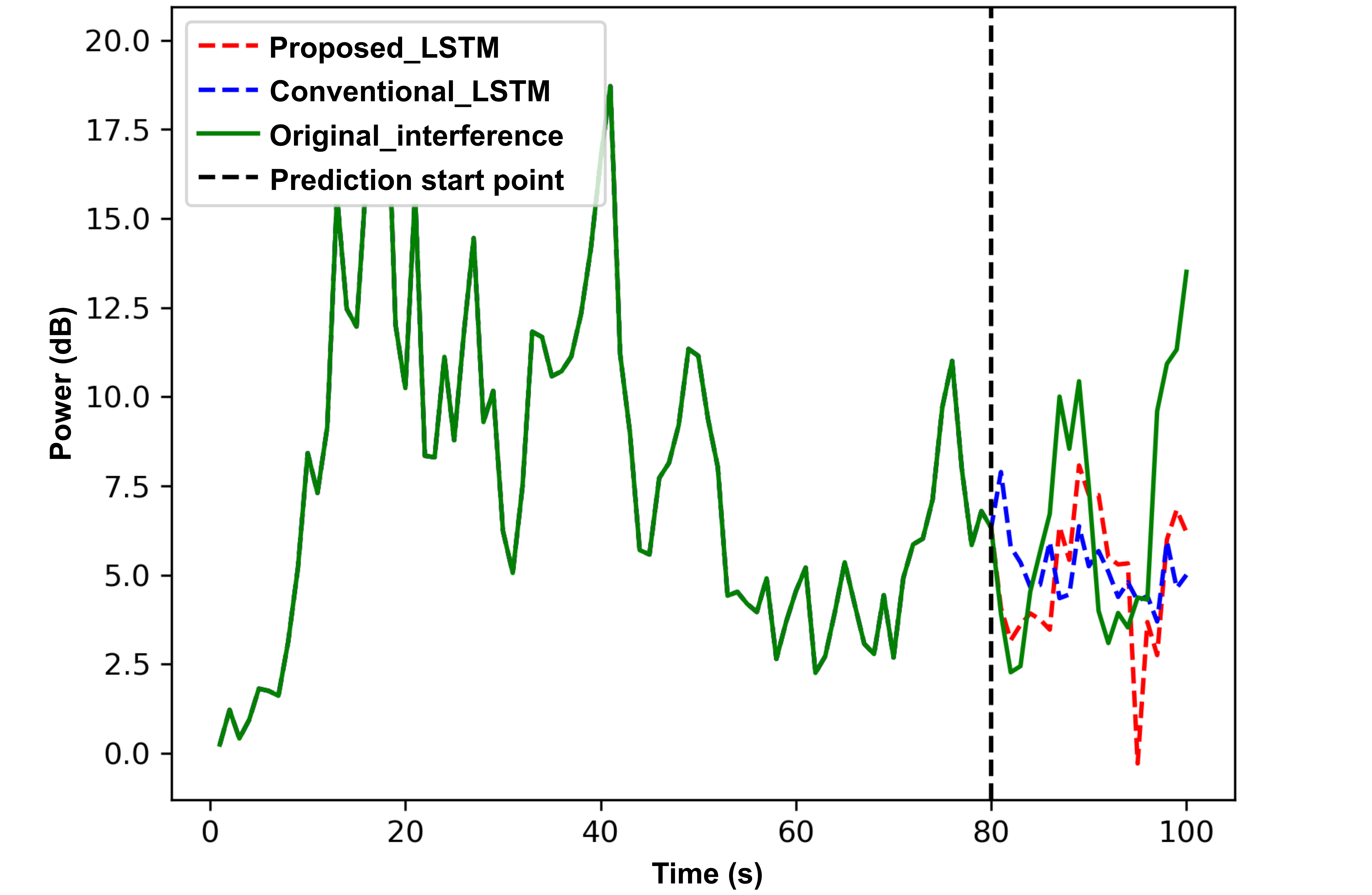}
    \caption{Comparison between proposed and conventional methods using LSTM model.}
    \label{fig:LSTM_comparison}
\end{figure}

\begin{figure}[ht]
    \center
    \includegraphics*[width=0.85\linewidth]{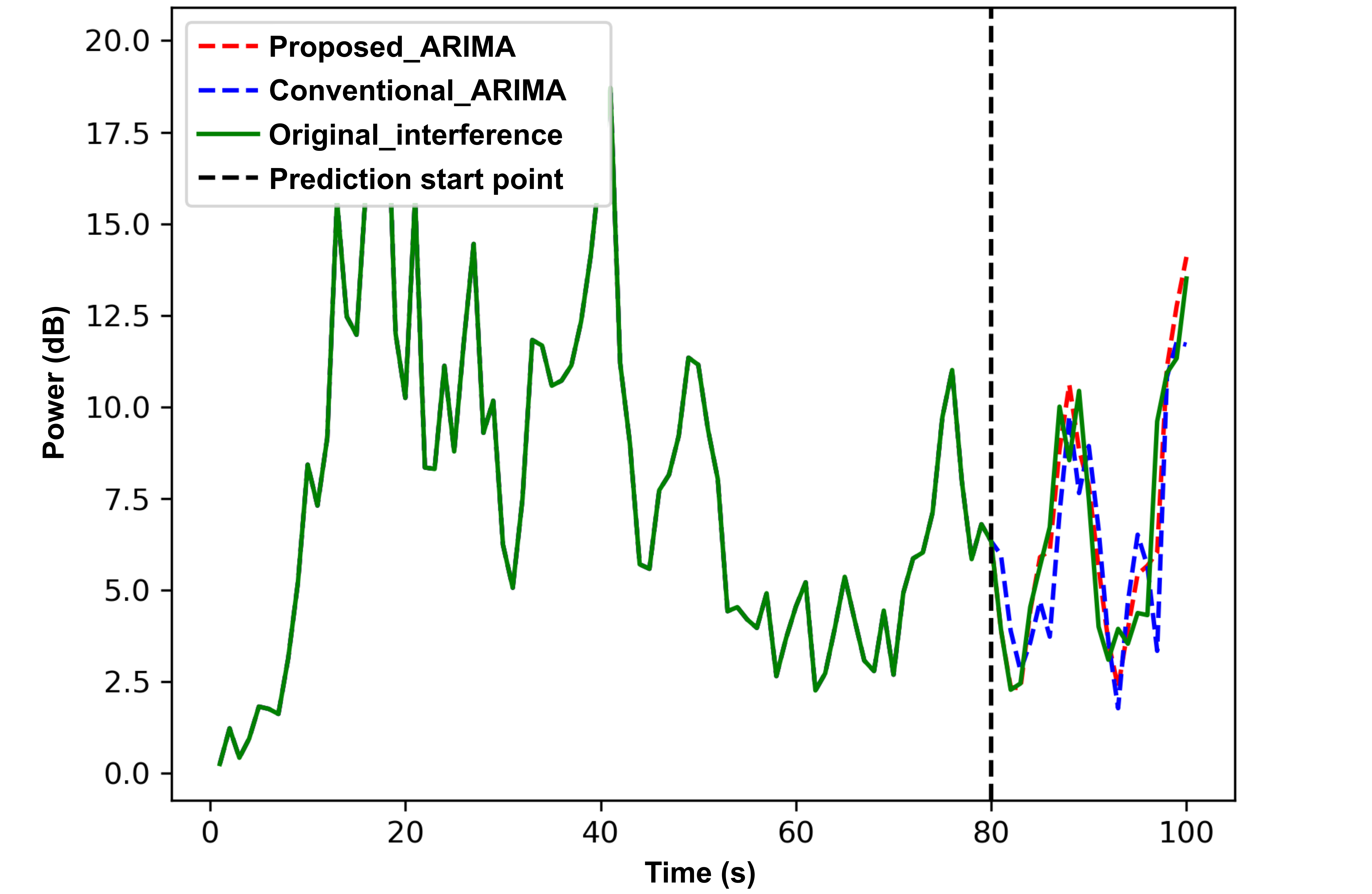}
    \caption{Comparison between proposed and conventional methods using ARIMA model.}
    \label{fig:ARIMA_comparison}
\end{figure}

Fig. \ref{fig:TF_comparison} illustrates the comparison between the conventional method and the proposed method for interference power forecasting using sequence-to-one transformer mode. The RMSE value of the conventional method and proposed decomposition-based method are $0.54$ and $0.77$, respectively. 

Fig. \ref{fig:LSTM_comparison} shows both the predicted interference signal power with the proposed and conventional approach results for LSTM. The RMSE value of the conventional method is $1.53$, whereas the RMSE of the proposed method is $1.71$. Fig. \ref{fig:ARIMA_comparison} shows both the predicted interference signal with the proposed and conventional approach results for ARIMA. For this, the RMSE of conventional and proposed methods are $0.57$ and $1.0$, respectively.

According to the results, it can be clearly seen that the transformer model outperforms both LSTM and ARIMA methods. The attention mechanism of the transformer architecture is capable of retaining and learning a considerable amount of past interference data is the main reason for this exceptional performance. Also, it is possible to justify this result as ARIMA and transformers perform well in lower mode IMFs while LSTM performs well in higher mode IMFs. However, the transformer model prediction is consistently well throughout all the decomposed components. By considering all the results obtained with the ARIMA, LSTM, and transformer model, Table \ref{tab4} summarizes the RMSE of the conventional and proposed methods for all the algorithms.

\begin{table}[htbp]
\caption{Summary of the RMSE values of conventional and proposed interference prediction methods.}
\begin{center}
\begin{tabular}{l c c}
\hline
\textbf{Algorithm}&\textbf{RMSE of conventional}&\textbf{RMSE of proposed} \\
&\textbf{method}&\textbf{method} \\
\hline

Transformers & \textbf{$0.77$} & \textbf{$0.54$}  \\
LSTM & $1.53$ & $1.71$ \\
ARIMA & $1.00$ & $0.57$ \\
\hline
\end{tabular}
\label{tab4}
\end{center}
\end{table}

\subsection{Resource Allocation}
\begin{figure}[htbp]
    \center
    \includegraphics*[width=0.85\linewidth]{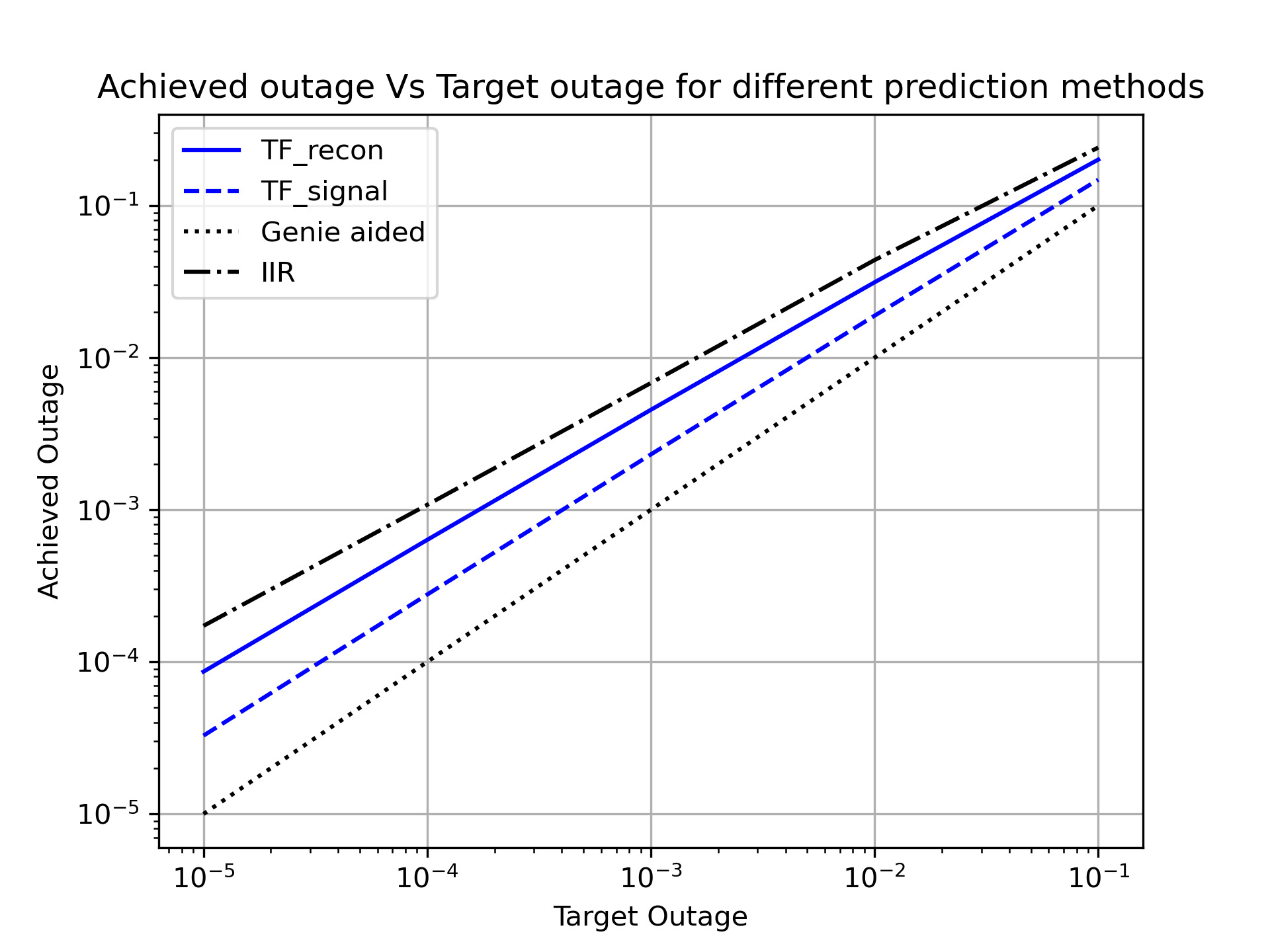}
    \caption{Variation of achieved outage with respect to target outage for transformers.}
    \label{fig:TF_outage}
\end{figure}

For the resource allocation task, as described in Table \ref{tab1}, $D = 50$ bits were taken as the communication bits, and the desired and interference signals were generated according to the specifications. Then, achieved target probability $\bar{\varepsilon}$ was calculated by considering different block error rate values $\varepsilon$. The resource allocation performance was evaluated by referring to Section\ref{Resource_allocation}. As the genie-aided scheme is the optimum resource allocation scheme since it has knowledge of the achieved SINR, it can be considered as the most efficient scheme. Fig. \ref{fig:TF_outage} illustrates the achieved outage concerning the target outage for the transformer model. It can be clearly seen that both conventional and proposed prediction methods based on outage curves are closely related to the genie-aided curve. Even though the EMD-based transformer model performs well in the interference prediction task, the conventional transformer method proves the best resource allocation efficiency compared to other methods. However, IIR filter-based resource allocation performs poorly than all the other methods.

\subsection{Interference Cancellation}
\begin{figure}[ht]
    \center
    \includegraphics*[width=1\linewidth]{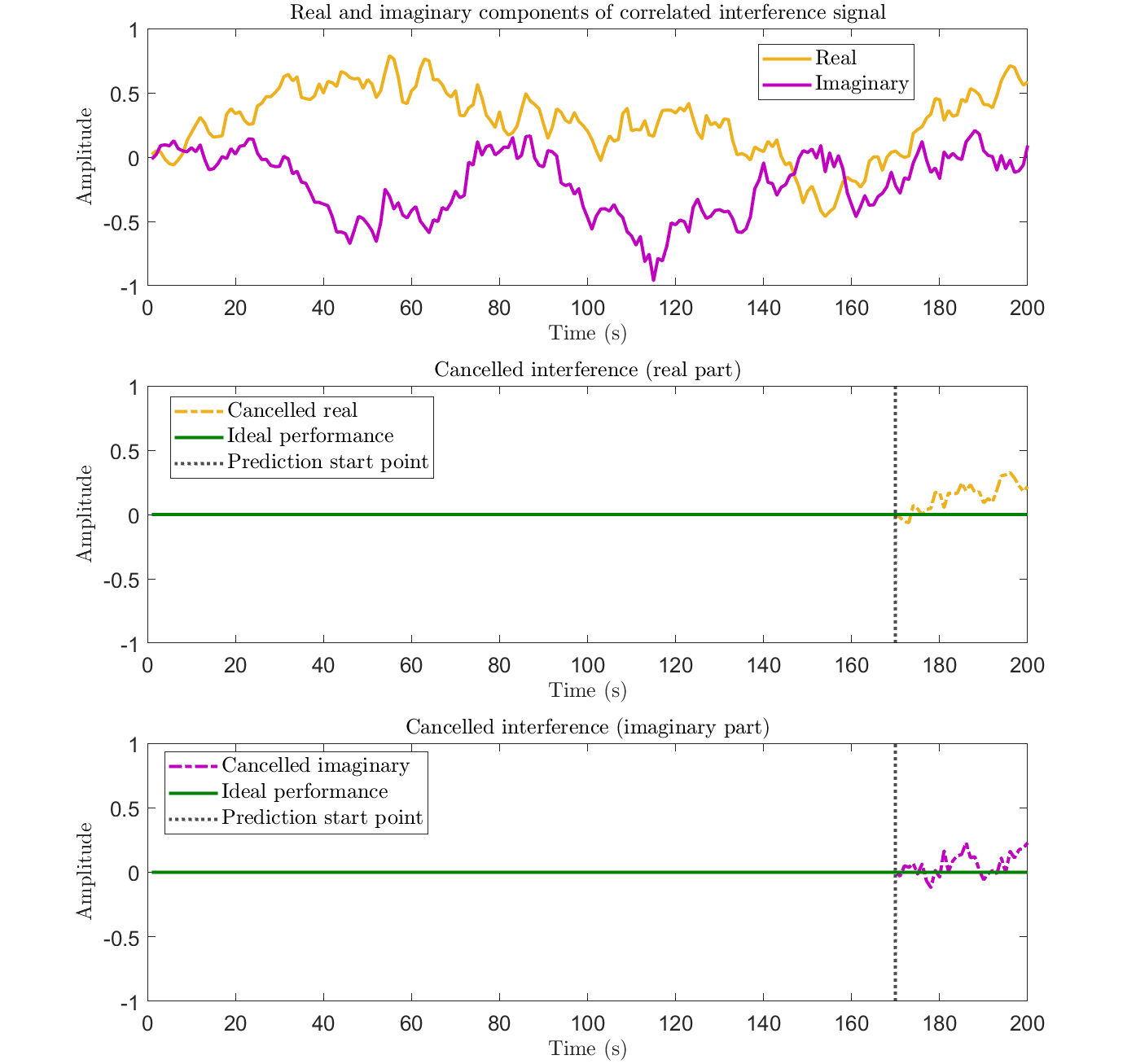}
    \caption{Interference cancellation based on the prediction using sequence-to-one transformer model.}
    \label{fig:Interference_cancellation}
\end{figure}

As mentioned in Section \ref{Interference_cancellation}, interference cancellation has been performed in the signal domain. Both the real and quadrature components were predicted using EMD based transformer method and canceled according to \eqref{eqn:real_cancellation} and \eqref{eqn:im_cancellation}. The first subplot of Fig. \ref{fig:Interference_cancellation} shows the real and quadrature components, whereas the second and third subplots compares the real and imaginary parts of the canceled interference signal against the ideal performance, respectively. %The RMSE values were calculated for each cancellation with respect to the ideal performance. Table \ref{tab5} summarizes the RMSE and SINR comparison.

% \begin{table}[htbp]
% \caption{RMSE and SINR comparison of interference cancellation.}
% \begin{center}
% \begin{tabular}{l c l c}
% \hline
% \textbf{Interference signal} & \textbf{RMSE} & \multicolumn{2}{c}{\textbf{SINR value (dB)}}\\
% \hline
% Real & 0.0685 & Before cancellation & 0.0082\\
% Quadrature  & 0.0423 & After cancellation & 2.3593\\
% \hline
% \end{tabular}
% \label{tab5}
% \end{center}
% \end{table}

\section{Conclusion}
\label{sec:conclusion}
In this paper, we have proposed a novel interference prediction scheme for local 6G networks. A sequence-to-one transformer algorithm accompanied by an advanced signal pre-processing technique known as EMD has mainly been utilized for the interference prediction task. In parallel, LSTM and ARIMA algorithms were used for comparison. In the conventional approach, the above three prediction models were applied to the original interference signal and obtained the predicted interference. For this method, the LSTM algorithm yielded the highest RMSE, while the sequence-to-one transformer model exhibited the lowest RMSE value. In the proposed approach, we have used EMD to decompose the original interference signal and apply three algorithms to predict decomposed signals separately. The total predicted interference signal was reconstructed by combining the predictions of individual IMFs and the residual. The transformer model exhibits the lowest RMSE for the proposed approach. Therefore the proposed sequence-to-one transformer model has yielded $54.97\%$ and $23\%$ gains against LSTM and ARIMA algorithms, respectively. We have explored resource allocation capability based on the predicted interference, and the conventional transformer model-based prediction has shown the best capability. Moreover, predicting the interference signal using the transformer algorithm may enable interference cancellation as an interference management approach. Finally, this study concluded that the proposed decomposition-based sequence-to-one transformer model provides promising results for wireless interference prediction and cancellation.

\bibliographystyle{IEEEtran}
\bibliography{main.bib}

\end{document}